\documentclass{article}
\usepackage{graphicx}
\usepackage{authblk}
\usepackage{amsmath,amssymb,amstext}
\usepackage{subfiles}
\usepackage{gensymb}
\usepackage[a4paper, total={7in, 10in}]{geometry}
\usepackage[colorlinks,linkcolor=blue,citecolor=blue,urlcolor=blue ]{hyperref}
\usepackage[export]{adjustbox}
\usepackage{subfig}
\usepackage{float}
\usepackage{multicol}
\usepackage{xcolor}
\usepackage{natbib}

\begin{document}
\title{Searching For JWST’s Little Red Dots}
\author{Jake Feeney$^{1,2}$}
\author{Patrick Kavanagh$^{1,2}$}
\author{John A. Regan$^{1,2}$}
\affil{$^1$Department of Physics, Maynooth University, Maynooth, Ireland}
\affil{$^2$Centre for Astrophysics and Space Science Maynooth, Maynooth University, Maynooth, Ireland}
\date{}
\maketitle

\setlength{\columnsep}{1cm}
\begin{abstract}
\noindent Recent observations by the James Webb Space Telescope (JWST) have revealed a previously hidden population of extremely bright and compact objects between redshifts of $z \sim 4$ and $z \sim 10$. Given their extreme red colouring in the observed frame these galaxies have been dubbed Little Red Dots (LRDs). The aim of this project was to identify LRDs using photometric data from previously uninvestigated JWST datasets and to estimate their AGN fractions by fitting the spectral energy distribution of each galaxy against well calibrated templates using CIGALE. We identified a list of potential LRDs using a single colour cut of F444W-F277W $>$1.5 mag and by applying a morphological analysis. We used EAZY to estimate the (photometric) redshift and CIGALE to estimate the AGN fraction of each LRD. Overall, we identified 14 LRDs, applying accurate SED fits to 11 of them. We found that 7 of them had a high AGN fraction (with the AGN component generating more than 50\% of the observed flux), a further two LRDs had AGN contribution in excess of approximately 40\%. In total nine LRDs (our of 14) are likely to have a supermassive black hole in their centre. Interestingly, the three LRDs which could not be well fit by CIGALE displayed extremely high photometric redshifts ($z_{phot} \gtrsim 11$) and require further analysis (and may also host a supermassive black hole in their centre).

\end{abstract}

\section{Introduction}
Due to the extreme sensitivity of the James Webb Space Telescope (JWST), observations have identified, previously undetectable, faint Active Galactic Nuclei (AGN) candidates at high redshift (\citealt{OgLRDmatthee2024littlereddotsabundant, Labb__2023,barro,AltCigaleParams}). One class of these newly identified AGN candidates are called Little Red Dots (LRDs) and they are defined by their ``v-shaped" spectral energy distribution (SED) (\citealt{kocevski2024risefaintredagn,OgLRDmatthee2024littlereddotsabundant,Labb__2023}).  There is an active debate as to what they are and more specifically what populations they host - be they dense stellar populations or active galactic nuclei. (\citealt{barro,akins2024cosmosweboverabundancephysicalnature,Inayoshi_2024}) For this reason more LRDs are being actively searched for and their underlying composition being thoroughly investigated.\\
\indent Given the numerous deep imaging observations with JWST, photometry and SED modelling can be used to interpret the stellar populations, stellar ages, AGN fractions and dust fractions on a large number of objects (\citealt{DPH,barro}). We are therefore reliant on photometric redshifts due to the relative lack of coverage of these regions with JWST's spectrographs. (\citealt{AltCigaleParams,kocevski2024risefaintredagn}).

\indent
In this work we searched for LRDs using JWST observations which were previously not used to search for LRDs. This work demonstrates that LRDs are ubiquitous in deep JWST imaging observations and that there is likely a large number yet to be identified in the publicly available data. A systematic survey of these data would likely increase the known population substantially and help resolve the nature of these objects.

\section{Data Analysis} \label{Section 2}
We searched the Mikulski Archive for Space Telescopes\footnote{\url{https://mast.stsci.edu/portal/Mashup/Clients/Mast/Portal.html}} (MAST) archive for JWST imaging surveys that used both the F444W and F277W filters alongside at least three additional filters with a minimum science duration of 5~ks. We chose four programmes to analyse in our study which are listed in Table~\ref{table:PIDs}. We reprocessed the data with the JWST Calibration Pipeline version 1.14.0 \citep{Bushouse_JWST_Calibration_Pipeline_2024} using versions 11.17.20 and `jwst\_1263.pmap' of the Calibration Reference Data System (CRDS) and CRDS context, respectively.

\begin{table}
\centering
\begin{tabular}{c p{0.5\linewidth} l }

\hline \hline
 PID  & Program Title & PI Name  \\ 
 \hline
 2756 & Imaging and Spectroscopic Follow-up of a Supernova at Redshift z=3.47 & Wenlei Chen  \\
 1199 & The metallicity of galaxies in the MACS J1149.5+2223 field & Massimo Stiavelli\\
 2767 & Imaging and Spectroscopy of Three Highly Magnified Images of a Supernova
at z=1.5 & Patrick Kelly\\
 1176 & JWST Medium-Deep Fields (Windhorst IDS GTO Program) & Rogier A. Windhorst \\
 \hline

\end{tabular}

\caption{The datasets we chose to use, based on their observation times and selection of filters.}
\label{table:PIDs}
\end{table}
\section{LRD Sample Selection}
We used the source catalogue with photometry produced by the \texttt{source\_catalog} step of the JWST Calibration Pipeline. We cross-matched sources in different filters using a source separation threshold of $0.2$ arcsec. We then applied the standard LRD colour cut of F444W-F277W $>$ 1.5 mag (\citealt{barro}). After that we removed all sources with less than five filters to ensure our SED modelling would be sufficiently constrained. Next we assessed the morphology of each of our candidates in each filter by eye to rule out extended sources as non-LRDs (see Fig.~\ref{Fig:Cigale Output}). Finally, we assessed the shape of the SEDs to eliminate everything that was not ``v-shaped" as described in \cite{kocevski2024risefaintredagn}. This resulted in 14 candidate LRDs across all datasets.
\section{SED Modelling}
\begin{figure}[H] 
\centering
\includegraphics[width=0.5\textwidth]{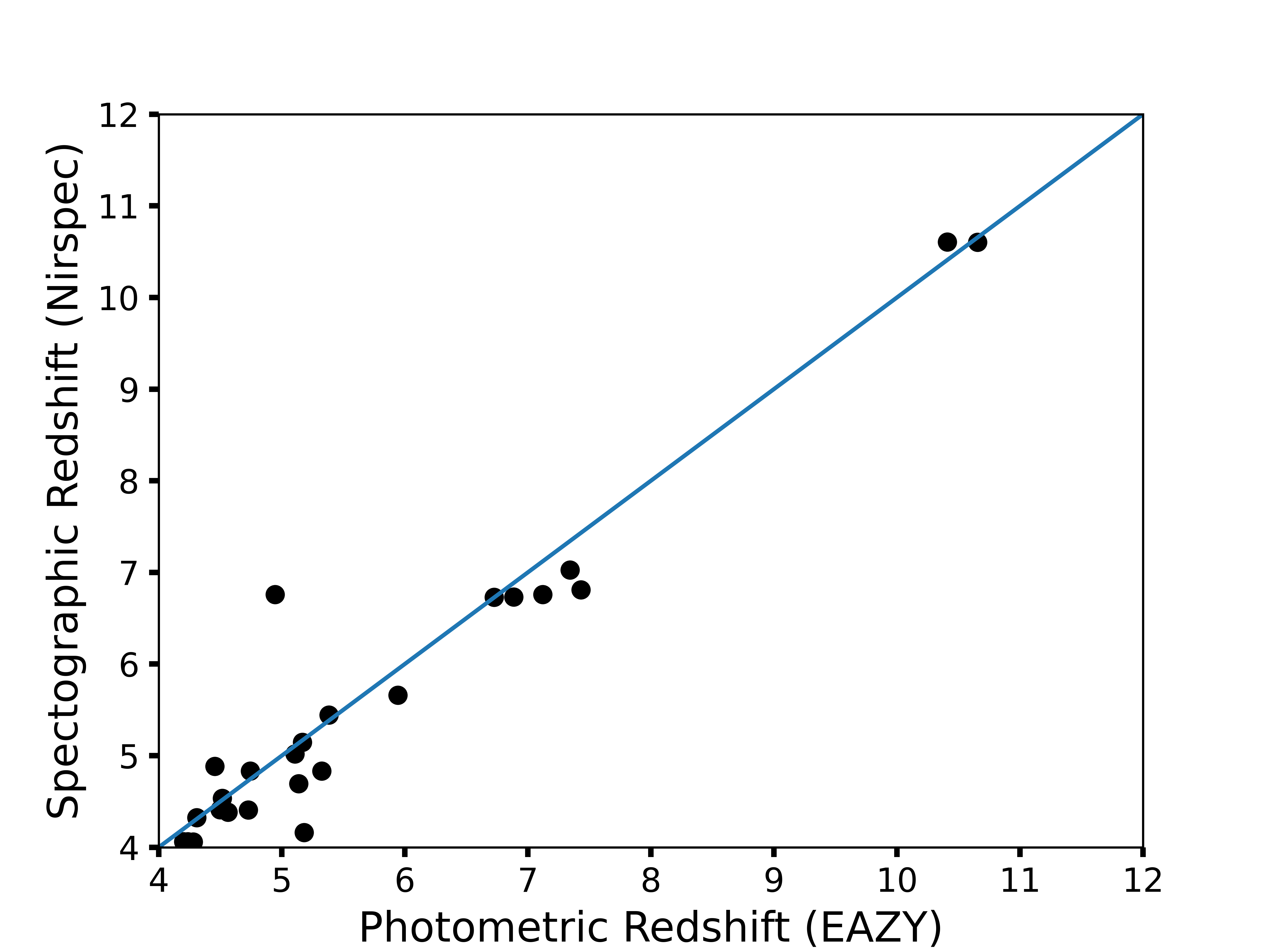}
\caption{Spectographic redshift vs photometric redshift from our bench-marking analysis using the JADES data. The photometric redshifts were determined using the software package EAZY \citep{EAZY}. The blue line is the $1:1$ fit. Overall the results from the photometric redshifts are in excellent agreement with those with known spectroscopic redshifts.}
\label{Fig:Figure1}
\end{figure}

\begin{figure}[H]
\centering
\subfloat{\includegraphics[width=0.475\textwidth, valign=c]{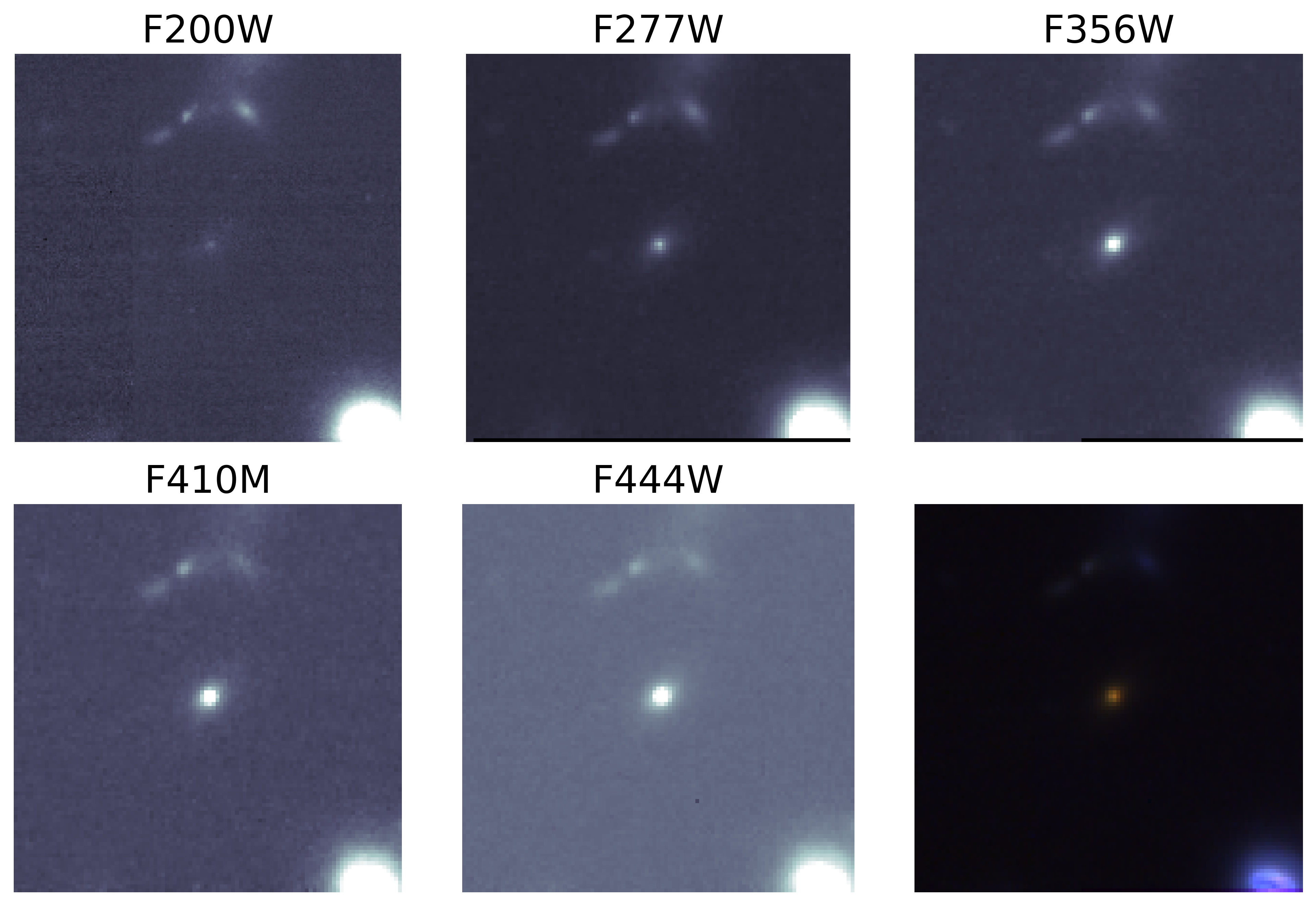}}
  \hfill
\subfloat{\includegraphics[width=0.475\textwidth, valign=c,trim={0 0 0 .65cm},clip]{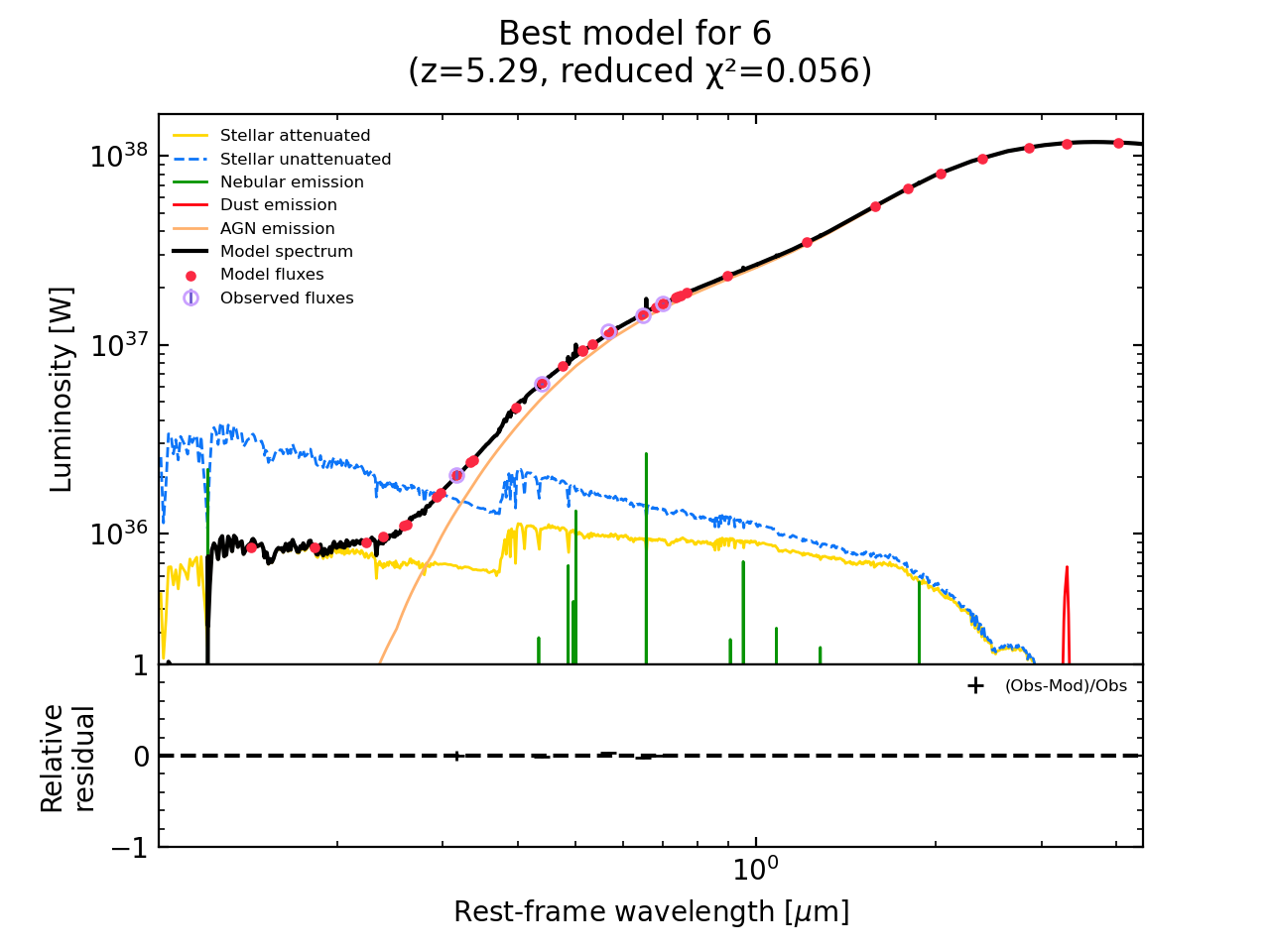}}
  \hfill
\subfloat{\includegraphics[width=0.475\textwidth, valign=c]{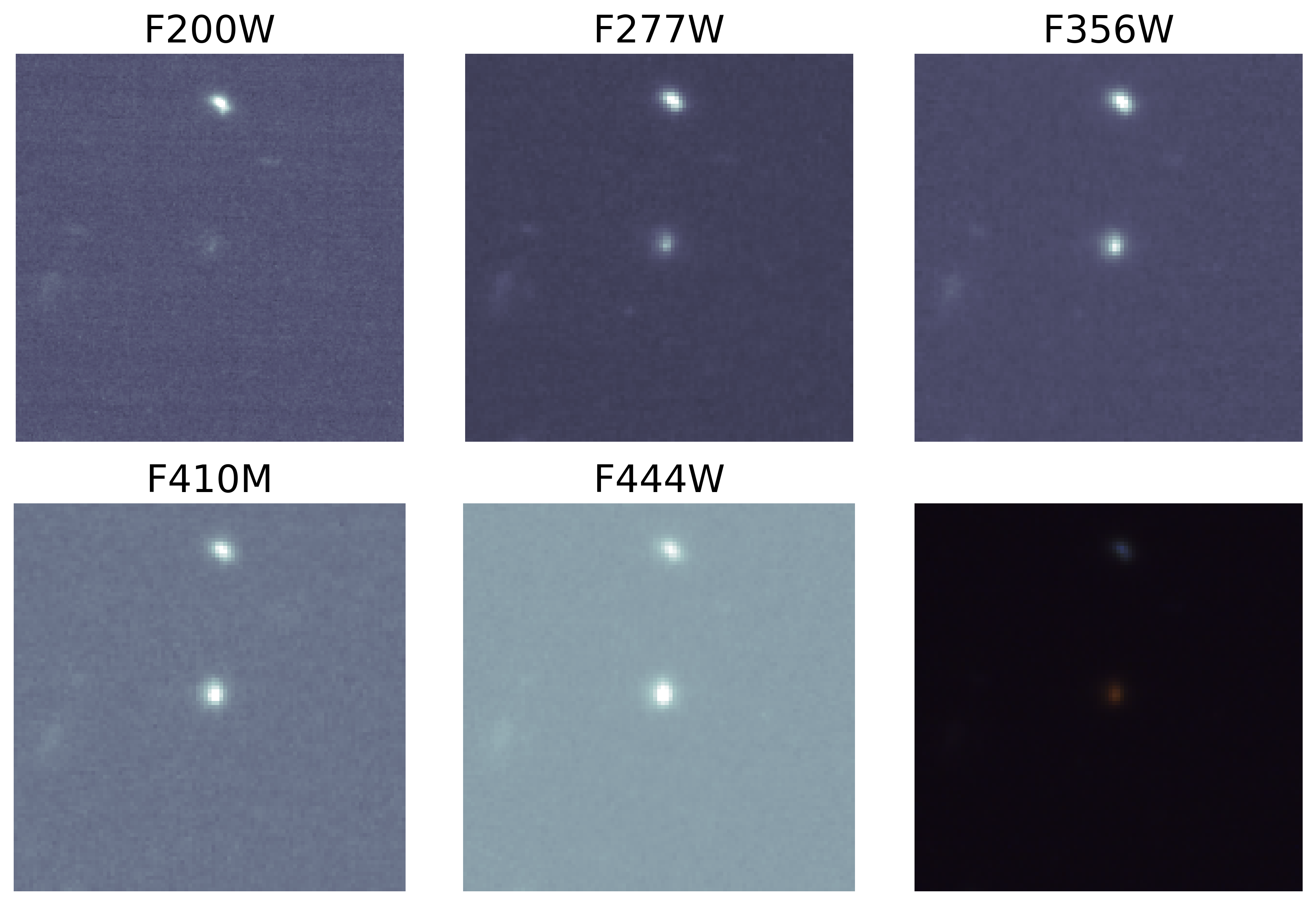}}
  \hfill
\subfloat{\includegraphics[width=0.475\textwidth, valign=c,trim={0 0 0 .65cm},clip]{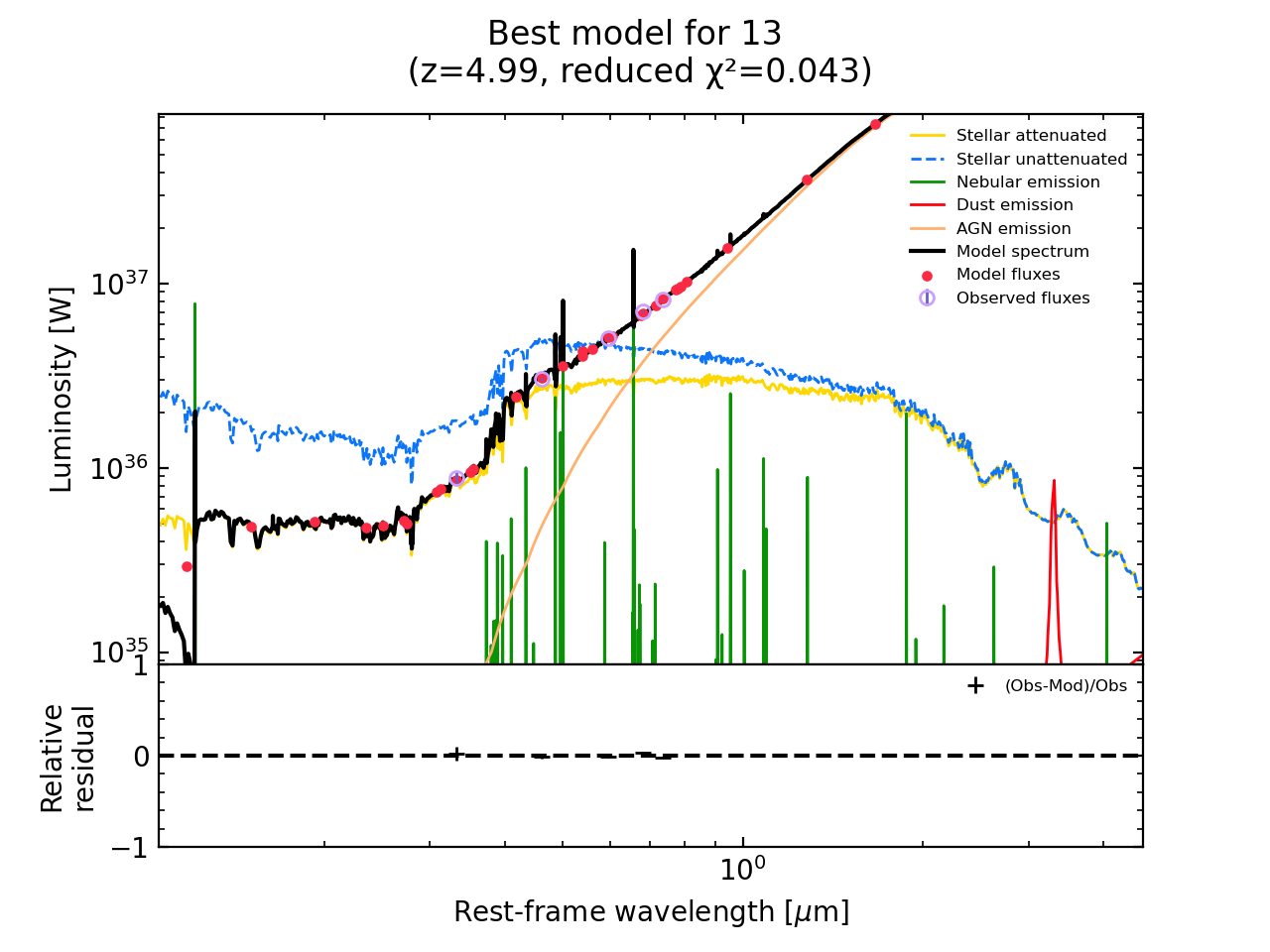}}
\caption{Left Panel: Photometry in five filters and the false colour images of two of our candidate LRDs. Right Panel: CIGALE best fit SED plots of the same two LRDs. In each case the LRDs are well fit by a stellar component and an AGN component. }
\label{Fig:Cigale Output}
\end{figure}

We used the EAZY package (\citealt{EAZY}) to obtain the photometric redshift for each of our candidate LRDs.
We configured it using the default template set “tweak fsps QSF 12 v3” which was created from the stellar population code FSPS (\citealt{Tweak}).
To determine the reliability of our photometric redshifts we bench-marked our methodology using JWST Advanced Deep Extragalactic Survey (JADES) (\citealt{eisenstein2023overviewjwstadvanceddeep, rieke2023jadesinitialdatarelease}) which provides the spectroscopic redshift for many sources in the survey field. We show the results of our photometric redshift determinations versus the known spectroscopic determinations in Fig.~\ref{Fig:Figure1} . Overall, the results are in excellent agreement with only one significant outlier at photometric redshift $z_{phot} \sim 5$.

\begin{figure}[H]
\centering

\subfloat[The AGN fraction of each LRD as a function of redshift. We see no obvious trend in our (small) sample.]{\includegraphics[width=0.313924051\textwidth,valign=c]{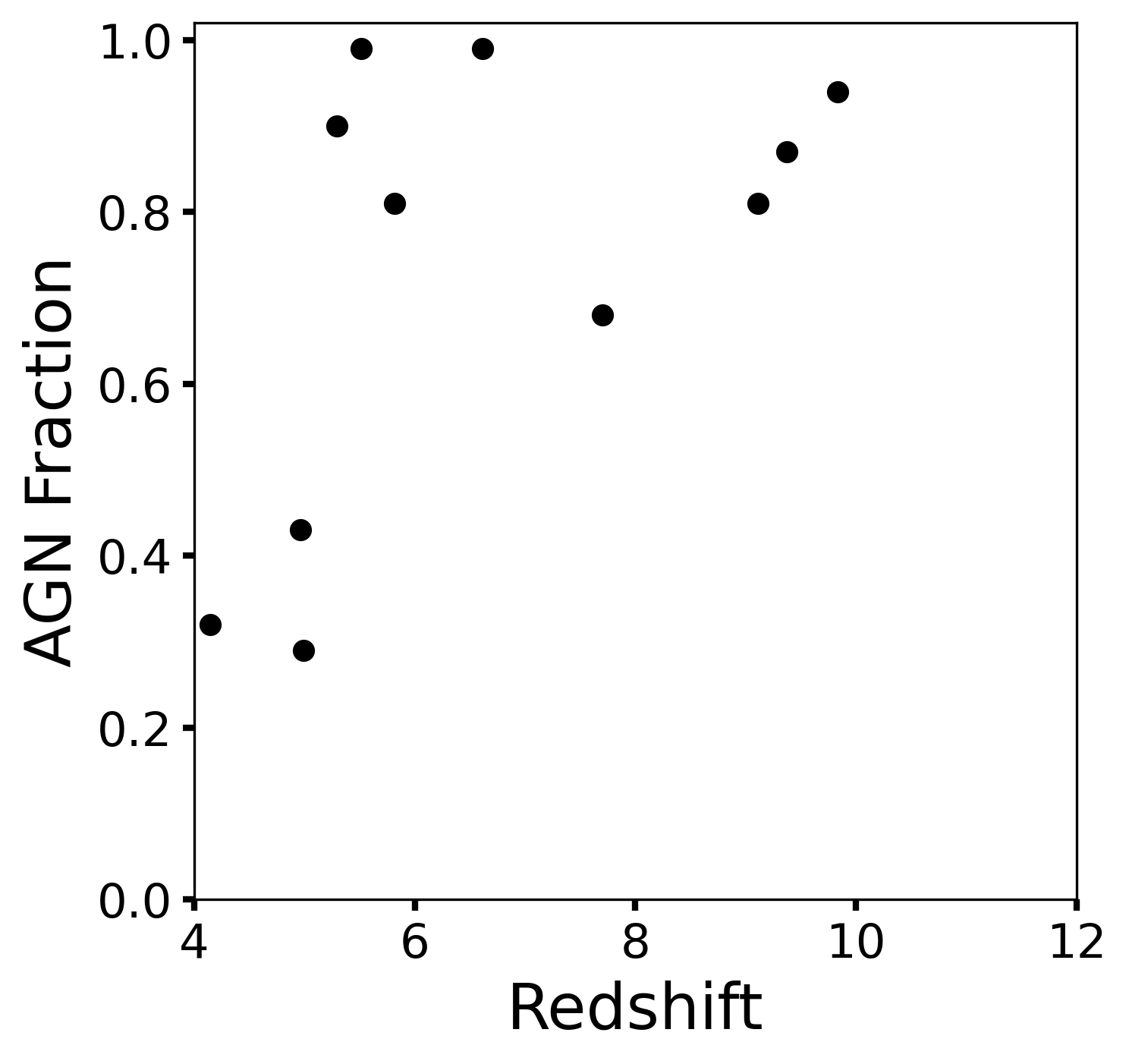}\label{fig:f1}}
  \hfill
\subfloat[The AGN Fraction distribution among the LRDs.]{\includegraphics[width=0.293037975\textwidth, valign=c]{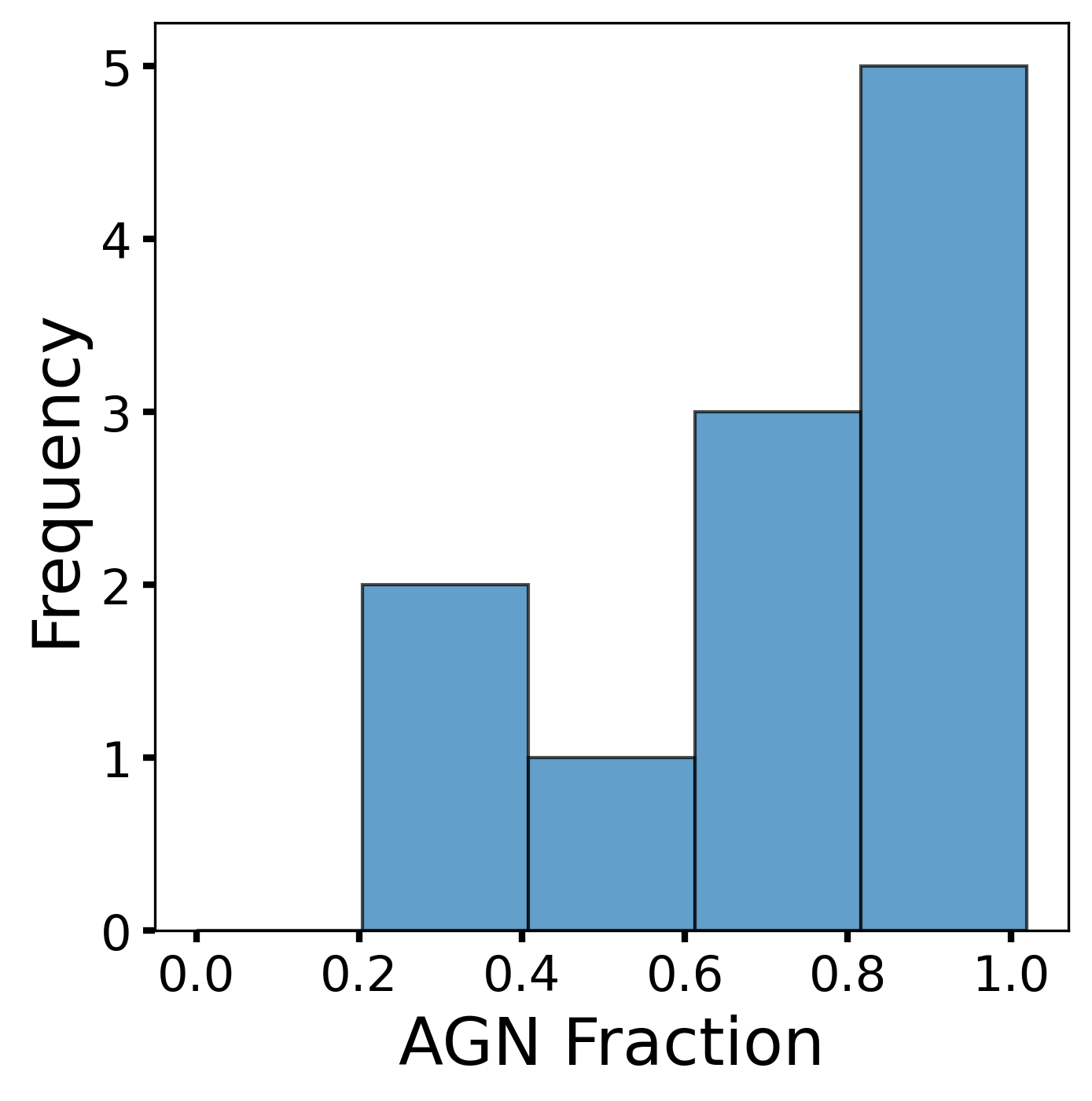}\label{fig:f2}}
  \hfill
\subfloat[The Redshift distribution among the LRDs.]{\includegraphics[width=0.293037975\textwidth, valign=c]{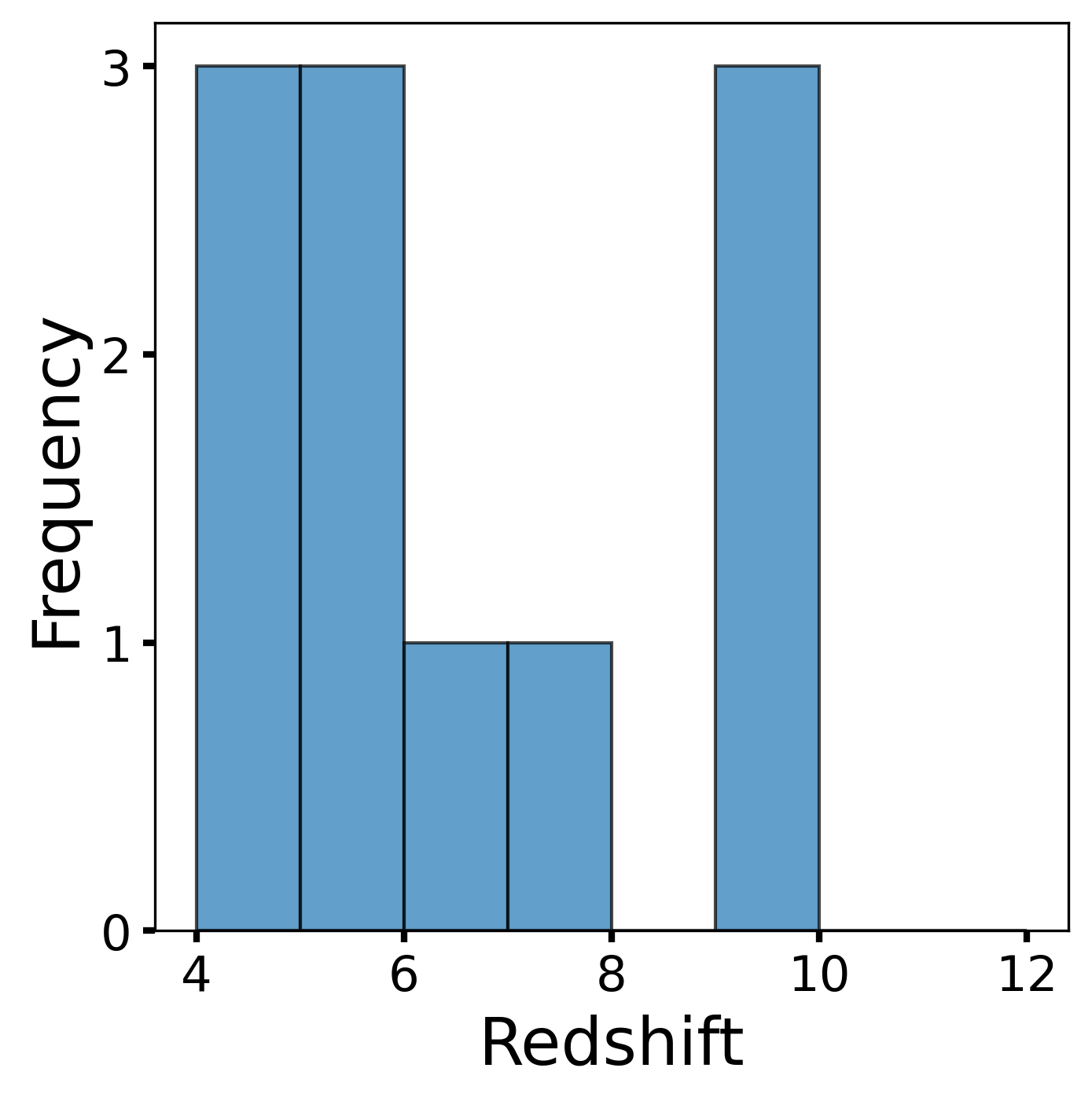}\label{fig:f3}}

\caption{Final Results from the SED Modelling. The LRDs that could not be
accurately SED modelled are omitted here.}
\label{Fig:Summary}
\end{figure}

\indent Having determined the photometric redshifts, we then used CIGALE (\citealt{CIGALE1,CIGALE2,CIGALE3}) to estimate the AGN contribution to the flux (or AGN fraction) in each LRD candidate. We used the modules \texttt{shfdelayed, bc03, nebular, dustatt\_modified\_starbust, dl2004, skirtor2016} and \texttt{redshifting} using parameters based on \citealt{DPH}. The only change from the \citealt{DPH} paper was that we added some higher values for the main stellar age in \texttt{sfhdelayed} (see Table \ref{table:Params} for a full list of parameters). We did this because some of the LRDs were higher redshift than the LRDs in \citealt{DPH}, so we raised the maximum value for stellar age to compensate.
We note that three of our candidate LRDs had $11 < z_{phot} < 12$ which could not be accurately fit using our modelling procedure - we therefore neglected these outliers in the final analysis and leave those for future analysis. Our final SED modelling results are in Table \ref{tab:Coords} and are visually represented in Figure \ref{Fig:Summary}. Six of the LRDs had redshifts between 4 and 6. The median redshift was 5.8163. Seven of the candidate LRDs have an AGN fraction greater than 0.8, implying that they have an AGN in their centre.

\begin{table*}[!htb]
\centering
\begin{tabular}{@{}l p{4.8cm} c c @{}}
\hline\hline
\textbf{Module} & \textbf{Parameter} & \textbf{Symbol} & \textbf{Values} \\ \hline

Star Formation History & e-folding time (main)  &  $\tau_{ main}$ &  0.1, 0.4, 0.8, 1, 2, 3, 4, 5 Gyrs \\  

\hspace{1cm} &Main stellar age &  $t_{star}$ &  0.1, 0.2, 0.4, 0.5, 0.6, 0.7, 1, 1.3, 1.6, 2 Gyrs \\

\hspace{1cm} &e-folding time (burst) &  $\tau_{burst}$ & 0.01, 0.05, 0.1 Gyrs \\

\fontfamily{cmtt}\selectfont{sfhdelayed} &Age of the burst population &  $t_{burst}$ & 0.01, 0.03, 0.05 Gyrs \\

\hspace{1cm} & Burst fraction &  $f_{burst}$ & 0, 0.001, 0.01, 0.1, 0.2, 0.3, 0.4, 0.5 Gyrs \\ \hline

Simple stellar population & Initial mass function & ... & Chabrier \\
\fontfamily{cmtt}\selectfont{bc03} & Mettalicity & $Z$ & 0.02 \\ \hline

Nebular Emission & Ionization parameter & log $U$ & -2.0 \\
\fontfamily{cmtt}\selectfont{nebular} & Gas Metallicity & $Z_{gas}$ & 0.02 \\ \hline

AGN emission & Optical depth at 9.7$\mu m$ & $\tau$ & 7 \\
\hspace{1cm} & Opening angle & $\theta$ & 40\degree \\
\hspace{1cm} & Viewing angle & $i$ & 30\degree \\
\fontfamily{cmtt}\selectfont{skirtor2016} & AGN contribution & frac$_{AGN}$ & 0.1 to 0.99 \\
\hspace{1cm} & Wavelength range for frac$_{AGN}$ & $\lambda_{AGN}$ & 0.53 $\mu$m\\ 
\hspace{1cm} & Extinction in polar direction & E(B-V) & 1, 2, 3, 4, 5, 6  \\ \hline

\end{tabular}
\caption{The parameters choices we used with CIGALE. All parameters not shown in this table are set to their default value as assigned by CIGALE.}
\label{table:Params}

\end{table*}

\begin{table}
\centering
\begin{tabular}{c c c c c }

\hline\hline
 PID  & RA (deg) & DEC (deg) & z$_{phot}$ & AGN Fraction \\ 
\hline
2756 & 3.592097 & -30.380469 & 5.5106 & 0.99 \\
2756 & 3.634261 & -30.437710 & 6.6114 & 0.99 \\
1199 & 177.375518 & 22.404280 & 7.7023 & 0.68 \\
1199 & 177.401306 & 22.429563 & 4.1407 & 0.32 \\
1199 & 177.4093204 & 22.465634 & 9.1109 & 0.81 \\
2767 & 322.396653 & 0.108467 & 11.3411 & - \\
1176 & 3.562758 & -30.390941 & 5.2929 & 0.90 \\
1176 & 15.702524 & -49.264881 & 5.8163 & 0.81 \\
1176 & 53.185419 & -27.680834 & 11.9051 & - \\
1176 & 53.187655 & -27.712561 & 11.5905 & - \\
1176 & 53.167202 & -27.715468 & 4.9898 & 0.29 \\
1176 & 64.043419 & -24.112786 & 4.9647 & 0.43 \\
1176 & 64.064852 & -24.096925 & 9.8365 & 0.94 \\
1176 & 171.796909 & 42.483895 & 9.3746 & 0.87 \\
\hline

\end{tabular}
\caption{The Program ID, coordinates (measured in F277W), redshift and AGN Fraction of our LRD candidates. The AGN Fraction of the candidates that could not be accurately modelled is not included.}
\label{tab:Coords}
\end{table}
\section{Summary}

To summarise, we have identified 14 candidate LRDs from the datasets listed in Table \ref{table:PIDs}. We calculated the photometric redshifts and AGN fraction for 11 of the candidate LRDs using a combination of EAZY and CIGALE. Our results are plotted in Figure \ref{Fig:Summary}. Seven of the candidate LRDs have an AGN contribution to the flux greater than 80\%. A further two LRDs require an AGN contribution greater than 40\%. Therefore nine of candidates are likely to have a supermassive black hole in their centre. Three of the LRD candidates could not be accurately modelled - possibly due to their extremely high (photometric) redshift ($z_{phot} > 11$). (See Table \ref{tab:Coords}) In this work we have found many LRDs in deep JWST imaging observations and therefore it is very likely that there are many unidentified LRDs in the publicly available JWST data.

\section*{Acknowledgements}
JF would also like to thank the Maynooth University Experiential Learning Department for running the SPUR program. JF would also like to thank  Dale Kocevski for helping him with EAZY. JR acknowledges support from the Royal Society and Science Foundation Ireland under grant number 
 URF\textbackslash R1\textbackslash 191132. JR also acknowledges support from the Irish Research Council Laureate programme under grant number IRCLA/2022/1165.

\bibliographystyle{apalike}
\bibliography{sample.bib}{}
\end{document}